\documentclass{iau}
\usepackage{graphicx,natbib}
 
\newcommand{\apj}{ApJ}           
\newcommand{\mnras}{MNRAS}       

\newcommand{\aap}{A\&A}

\newcommand{\aj}{AJ}

\newcommand{\vmax}{\mbox{$V_{\rm max}$}}

\title{Kinematic Evolution of Field and Cluster Spiral Galaxies}

\author[Ziegler \& B\"ohm]{Bodo L. Ziegler$^1$ \and Asmus B\"ohm$^2$}

\affiliation{$^1$University of Vienna, Department of Astrophysics, T\"urkenschanzstr. 17, 1180 Vienna, Austria \\
email: {\tt bodo.ziegler@univie.ac.at}\\
$^2$Institute for Astro- and Particle Physics,
Technikerstrasse 25/8,
6020 Innsbruck,
Austria 
email: {\tt asmus.boehm@uibk.ac.at}}

\pubyear{2014}
\volume{311}
\jname{Galaxy Masses as Constraints of Formation Models}
\editors{M. Cappellari \& S. Courteau, eds.}

\begin{document}

\maketitle

\begin{abstract}
We investigate the evolution of the Tully--Fisher relation out to $z=1$ with 137 emission-line galaxies in
the field that display a regular rotation curve. They follow a linear trend with lookback time being on
average brighter by 1.1\,$B$mag and 60\% smaller at $z=1$. For a subsample of 48 objects with very regular
gas kinematics and stellar structure we derive a TF scatter of 1.15mag, which is two times larger than
local samples exhibit. This is probably due to modest variations in their star formation history and chemical
enrichment.

In another study of 96 members of Abell\,901/902 at $z=0.17$ and 86 field galaxies with similar redshifts
we find a difference in the TFR of 0.42mag in the $B$-band but no significant difference in stellar mass.
Comparing specifically red spirals with blue ones in the cluster, the former are fainter on
average by 0.35\,$B$mag and have 15\% lower stellar masses.
This is probably due to star formation
quenching caused by ram-pressure in the cluster environment. Evidence for this scenario comes from
strong distortions of the gas disk of red spirals that have at the same time a very regular stellar disk structure.
\end{abstract}

\firstsection
\section{Introduction}

Internal kinematics of galaxies provide important clues about their nature and reveal physically distinct
components
\citep[e.g.][]{k14}.
While stars are on colissionless orbits tracing the gravitational potential dominated by dark matter,
gas clouds can react faster on subtle effects and may display peculiar motions revealing interaction events.
In case of undisturbed spiral galaxies rotation produces a characteristic velocity profile with a steep
rise in the inner part turning over to a flat part, the so-called rotation curve RC.
In such a virialized state, the maximum velocity \vmax\ provides a good estimate of the total mass of a
galaxy and tightly correlates with the luminosity in the Tully--Fisher relation TFR.
Under this assumption the TFR is an important scaling relation to probe quantitatively the evolution of
disk galaxies with redshift 
as applied in several studies before
\citep[e.g.][]{z02,bz07,m11,t11}.
Since stellar kinematics from absorption lines requires a higher $S/N$ most studies of distant galaxies
rely on emission line measurements from the warm ionized gas.
Here, we report on two applications: a TFR study of distant field galaxies and an investigation of 
transition objects in the cluster complex Abell\,901/2.

\section{Evolution of the Tully--Fisher relation to $\mathbf{z=1}$}

\begin{figure}[t]
\centering
\includegraphics[angle=-90,width=0.7\columnwidth]{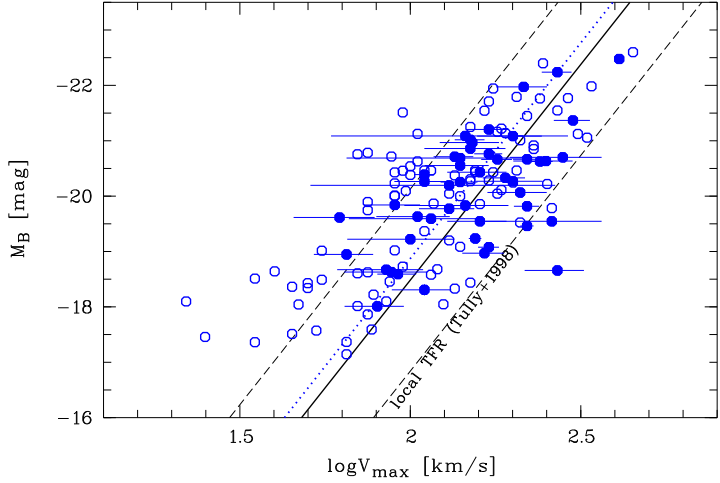} 
\caption{
The Tully--Fisher relation of 137 field galaxies with $0.05<z<0.97$ (open symbols) compared to the
local relation from Tully et al. 1998 (solid line \& 1-$\sigma$ boundaries dashed).
Filled circles represent a subsample of 48 disk galaxies with low morphological asymmetry and high
quality rotation curves.
Fitting their zeropoint yields the dotted line offset by 0.40mag towards brighter luminosities.
}\label{figtf}
\end{figure}

\begin{figure}[b]
\centering
\includegraphics[angle=-90,width=0.7\columnwidth]{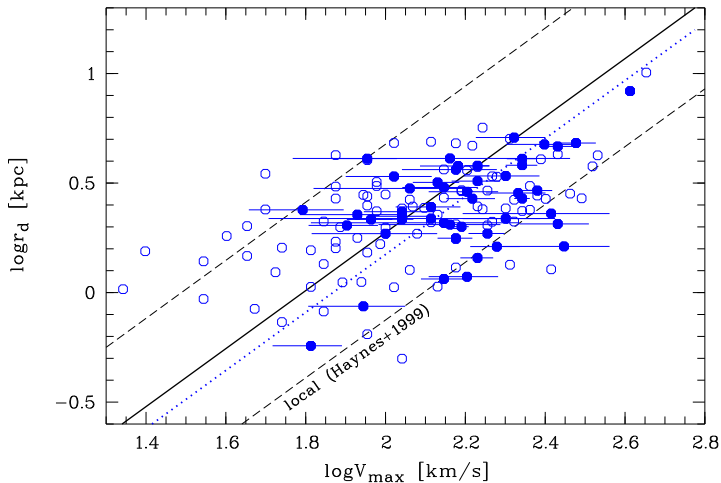} 
\caption{
The Velocity--Size relation of the same 137 galaxies like in Fig.\,\ref{figtf} 
this time compared to the local relation from Haynes et al. (1999).
On average, distant galaxies have disk scale lengths $r_d$ smaller by
0.15\,dex than local ones for given \vmax\ (fit indicated by dotted line).
}\label{figvsr}
\end{figure}

Continuing our previous investigations
we expanded our sample towards fainter galaxies with new observations 
with ESO's VLT using FORS
\citep{bz14}.
The whole sample holds high-quality spectra of 238 objects.
For 137 disk galaxies we were able to extract RCs and determine \vmax, while 101 emission-line
objects were rejected from our TF analysis due to kinematic distortions.
These field spirals are spread like $0.05<z<0.97$ with a median redshift of 0.45 corresponding to
lookback times between 0.6 and 7.6\,Gyr and $\langle t_{\rm lb}\rangle=4.5$\,Gyr.
In Fig.\,\ref{figtf} we show their TFR in rest-frame $B$-band compared to the local study of \cite{t98}.
The slope of the distant sample is with $-7.16^{+0.71}_{-0.53}$ similar to the local value of $-7.79$.
We further create a subsample of 48 objects restricted to have both high-quality RCs and a low
structural asymmetry 
($A_{\rm morph}\!<\!0.25$ in the CAS system of \cite{c00}) to ensure regularity in both the gas and
the stellar component in order to avoid any non-relaxed contribution to the velocity field.
Assuming the local slope this subsample shows moderate brightening of on average
$\langle \Delta M_B \rangle (z\!\approx\!0.5) = -0.40^m$
and a scatter of $1.15^m$, which is two times larger than the local one.
This may be attributed to a modest diversity of star formation histories 
and chemical enrichments
of the observed galaxies
\citep{fbzs14}.
Looking at the evolution of the TF residuals from the local relation we fit our whole sample linearly with
lookback time and find 
$\Delta M_B = -(3.63\pm1.78) \log(1+z)$
albeit with large scatter.
Such an increase in luminosity is derived by other studies, too \citep[e.g.][]{m11},
and in accordance with model predictions \citep{d11}.

We also investigate the Velocity--Size evolution of the same galaxies by measuring their disk scale lengths $r_d$ in HST $I$-band images.
Although the scatter is rather large, our sample is offset by $-$0.15\,dex 
to smaller $r_d$ for given \vmax\ compared to the
local sample of \cite{h99}, see Fig.\,\ref{figvsr}.
Note that the local sample was observed in $I$-band, too, and the scale
lengths of the distant galaxies were slightly corrected for rest-frame
wavelength shift following \cite{dejo96}.
On average, disk sizes decrease by
$\Delta \log r_d = -(0.59\pm0.38) \log(1+z)$
becoming about 60\% smaller at $z=1$, which is again in accordance with models.

\section{Red spirals as stripped transition objects in clusters}

Looking for environmental effects  we examine galaxies in
the multiple cluster system Abell\,901/902 at $z=0.17$ \citep{gray09}.
An observational program with ESO's VLT using VIMOS targeted 
$\sim$\,200 emission-line galaxies in the cluster
complex with tilted slits placed along the major axis
(PI A.~B\"ohm, see \cite{bb13a}). 
A kinematic analysis was possible for 96 cluster members and 86 field spirals with a similar average redshift 
($0.12<z<0.38$).
However, many objects don't exhibit a regular RC but distorted kinematics.
To quantify the degree of distortion we introduce similar to \cite{d01}
an RC asymmetry index $A_{\rm RC}$ based on differences of the approaching and receding arms.
We find an average value of $\langle A_{\rm RC} \rangle = 17\%$ in the field,
while  distortions are more frequent in the cluster, yielding
$\langle A_{\rm RC} \rangle = 25\%$.
Restricting to objects with high-quality RCs
for a TF analysis we find that cluster members are
on average modestly fainter in rest-frame $B$-band by 
$\Delta M_B  = 0.42\pm0.15^m$.
However, there is no significant difference in the stellar TF with
$\Delta \log M_* = 0.00\pm0.07$.

\begin{figure}[t]
\centering
\includegraphics[angle=-90,width=0.75\columnwidth]{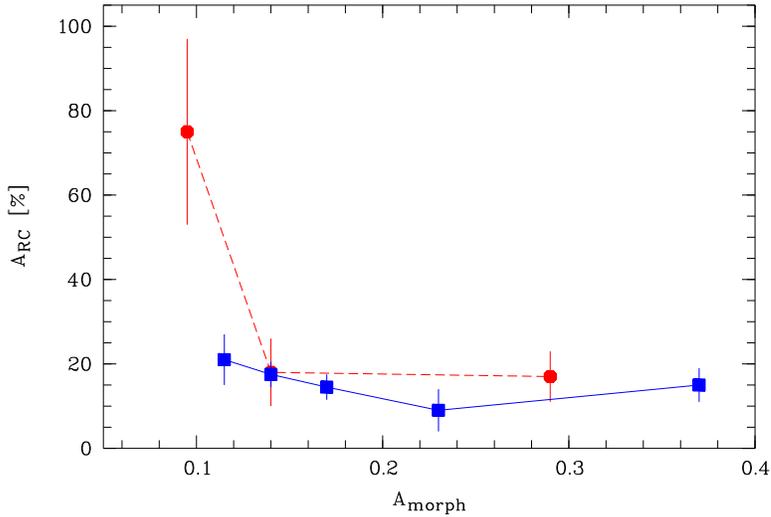} 
\caption{
Rotation curve asymmetry $A_{\rm RC}$ versus morphological asymmetry 
$A_{\rm morph}$
for red spirals
(circles) and blue spirals (squares) in the cluster system A901/902.
Average error-weighted values have been computed for bins holding 9
(red) and 12 objects (blue), respectively.
A large fraction of red spirals show strong rotation curve asymmetries, 
i.e.~disturbed gas kinematics, combined with undistorted stellar structure (i.e.~low
$A_{\rm morph}$). This can best be explained by ram pressure due to the intra-cluster medium.
}
\label{amarc}
\end{figure}

Further, we concentrate on red spirals that were defined by \cite{w09} to have red colors similar to red-sequence galaxies but 
(spiral) disk morphology and significant star formation rates (SFRs), 
albeit lower than typical for blue spirals.
Those red spirals with regular RCs are on average fainter in the $B$-band TFR by 
$\Delta M_B  = 0.35\pm0.12^m$ and show a slight difference in the stellar TF by
$\Delta \log M_* = 0.06\pm0.05$
compared to blue-cloud cluster members
\citep{bb13b}.
These results may be explained by a recent quenching of SF for red spirals caused by ram-pressure stripping
in the cluster removing gas.
This interpretation is strengthened by 
Fig.\,\ref{amarc}: many red spirals show strong rotation curve asymmetries
combined with an undisturbed stellar morphology. The only
cluster-specific interaction process which affects the gas disk but not the
stellar morphology is ram-pressure \citep{kron08}.
Thus, our analysis adds evidence that red spirals are transition objects between field blue-cloud and cluster lenticular galaxies.

\section*{Acknowledgements}

\noindent
Based on ESO PID 081.B-0107, 384.A-0813 and other programs. AB acknowledges
support by the  Austrian Science Foundation FWF (grants
P19300-N16 and P23946-N16). 
BZ thanks SOC and LOC for a very nice symposium.

\end{document}